\documentstyle{article}
\include{dafne.sty}
\newcommand{\beq}{\begin{equation}}
\newcommand{\eeq}{\end{equation}}
\newcommand{\eq}[1]{(\ref{#1})}

\newcommand{\beqn}{\begin{eqnarray}}
\newcommand{\eeqn}{\end{eqnarray}}

\newcommand{\e}{\mbox{${\bf e}$}}

\newcommand{\pch}{\mbox{$2\pi^+2\pi^-$}}
\newcommand{\pne}{\mbox{$\pi^+\pi^-2\pi^0$}}
\newcommand{\epm}{\mbox{$e^+e^-$}}

\begin{document}
\title{ 
$a_1\,\pi$ CONTRIBUTION TO $e^+e^-\to 4\pi$ ANNIHILATION AND
$\tau\to 4\pi\nu_{\tau}$ DECAY}
\author{
A.I.~Milstein   \\
{\em   Budker Institute of Nuclear Physics, Novosibirsk, 630090, Russia
}} 
\maketitle
\baselineskip=11.6pt
\begin{abstract}
The results of the study of the process $\epm \to 4\pi$
by the CMD-2 collaboration at VEPP-2M are presented. 
Analysis of the differential distributions demonstrates the
dominance of the $a_1 \pi$ and $\omega \pi$ intermediate states.
Simple model,
based on the assumption of $a_1(1260)\pi$ and
$\omega\pi$ dominance as intermediate states,
successfully  describes also the data of CLEOII and ALEPH
obtained recently for the decay $\tau\to 4\pi\nu_{\tau}$.
\end{abstract}
\baselineskip=14pt
\section{Introduction}
Production of four pions is one of the dominant processes of
$\epm$ annihilation into hadrons in the energy range from 1.05 to 2.5 GeV.
Due to the conservation of vector current (CVC) the cross section of
this process is related to the probability of $\tau\to
4\pi\nu_{\tau}$ decay \cite{CVC1}. Therefore, all realistic models describing
first process should also be appropriate for description of another one.

One of the main difficulties in the experimental studies of
four pion production is caused by the existence of
different intermediate states via which the final state could be
produced, such as
$\omega\pi$, $\rho \sigma$,
$a_{1}(1260) \pi$, $h_{1}(1170) \pi$, $\rho^{+} \rho^{-}$,
$a_{2}(1320) \pi$, $\pi(1300) \pi $.
The abundance of various possible mechanisms and their
complicated interference results in the necessity of simultaneous
analysis of two possible final states ($\pch$ and $\pne$).

In Section II of this report I present results from a
model-dependent analysis of both possible channels in $\epm$
annihilation into four pions based on data collected with  the CMD-2 detector
in the energy range 1.05-1.38 GeV  \cite{our}. 
To describe four pion production a simple model was used assuming
quasitwoparticle intermediate states and taking into account 
the important effects of the identity of the final pions 
as well as the interference of all possible amplitudes. 
It was unambiguously demonstrated that the main
contribution to the cross section  of the process $\epm\to 4\pi$
in the energy range 1.05 -- 1.38 GeV  , in addition to previously well-studied
$\omega\pi^0$,
is given by  $\rho\pi\pi$ intermediate state.
Moreover, the latter is completely saturated by the $a_1\pi$ mechanism.
The contribution of other intermediate states was
estimated  to be less than 15 \% .

In Section II I also discuss the results for  $\epm \to \pch$ cross
section obtained recently  by CMD-2 in the energy region 0.60--0.97
GeV \cite{our2}. 
In this energy region the energy dependence of the cross section
agrees with the assumption of the $a_1\pi$ intermediate state which is
dominant above 1 GeV. 


In Section III  the comparison is performed of the available
experimental data for   $\tau\to 4\pi\nu_{\tau}$ decay
\cite{Argus,Aleph,Cleo1,Cleo2} with the prediction of the model based on
the assumption of the $a_1\pi$ and $\omega\pi$ dominance \cite{our1}. 
It is shown that the model successfully  describes experimental data on
$\tau\to 4\pi\nu_{\tau}$ decay.

\section{The process $e^+e^-\to 4\pi$ }
The results described here are based on 5.8 pb$^{-1}$ of $\epm$
data collected at energies 2$E_{beam}$ from 1.05 up to
1.38 GeV at the VEPP-2M collider with the CMD-2 detector \cite{our}.

\begin{figure}[ht]
\vspace{9.0cm}
\includegraphics{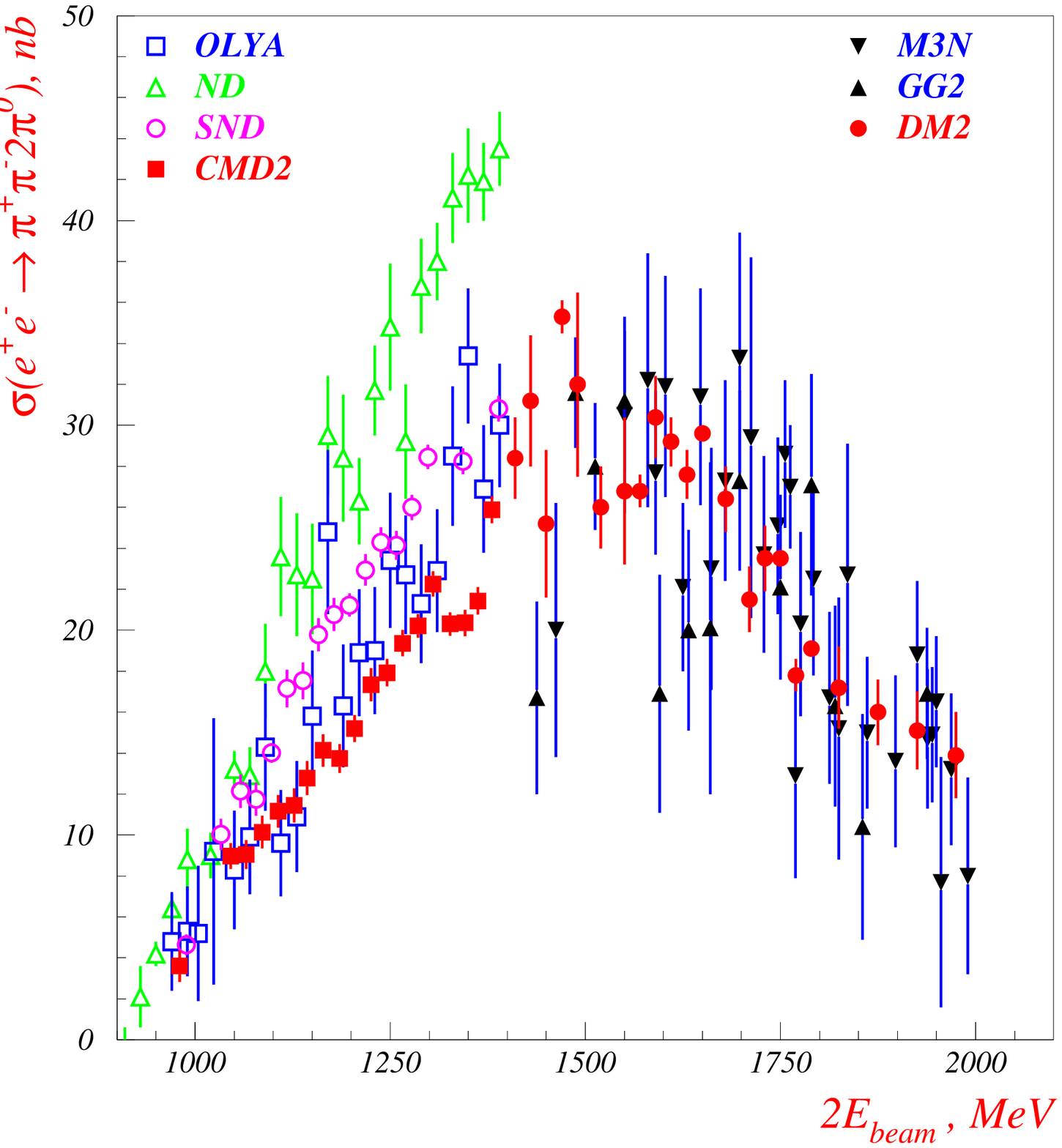}
\caption{\it
Energy dependence of the $\pne$
cross section}
\label{xs2pi}
\end{figure}

\begin{figure}[ht]
\vspace{9.0cm}
\includegraphics{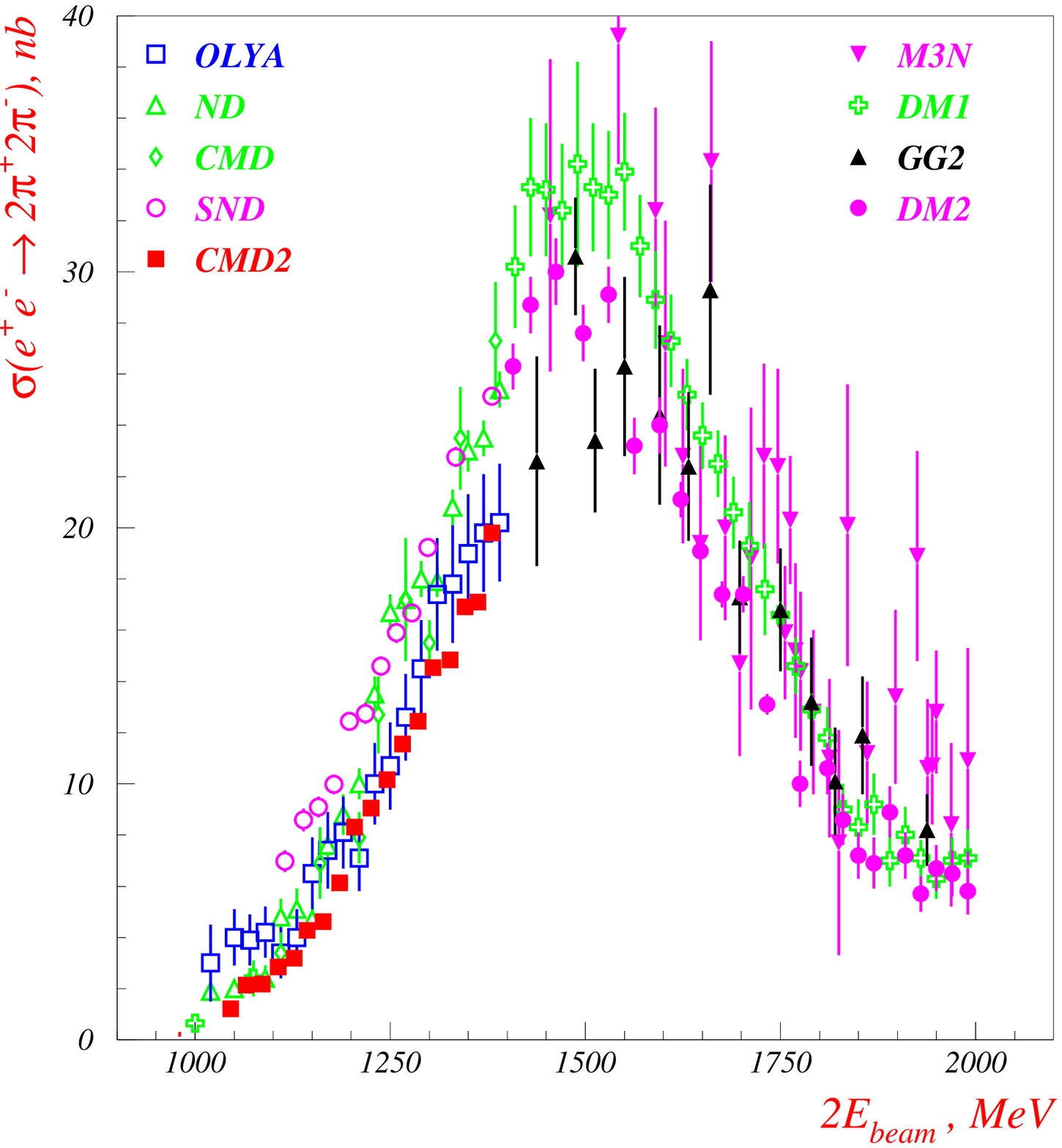}
\caption{\it
Energy dependence of the $\pch$
cross section}
\label{xs4pi}
\end{figure}

Figures~\ref{xs2pi},~\ref{xs4pi} show the total cross
sections  $e^+e^-\to \pne$ and $e^+e^-\to \pch$ vs energy. 

The cross sections measured in \cite{our} is consistent with the 
previous measurement at OLYA \cite{olya1},  CMD \cite{cmd},  and
within systematic errors don't  
contradict to the recent result from SND \cite{snd98}. For  $e^+e^-\to \pch$
the value of the cross section from all three groups is significantly lower
than that from ND \cite{nd,ndr}. Above 1.4 GeV  the
results from Orsay \cite{dm1,dm2,m3n} and Frascati \cite{mea,gg2:xs,gg2}
groups are shown. 

\begin{figure}[ht]
 \vspace{9.0cm}
\includegraphics{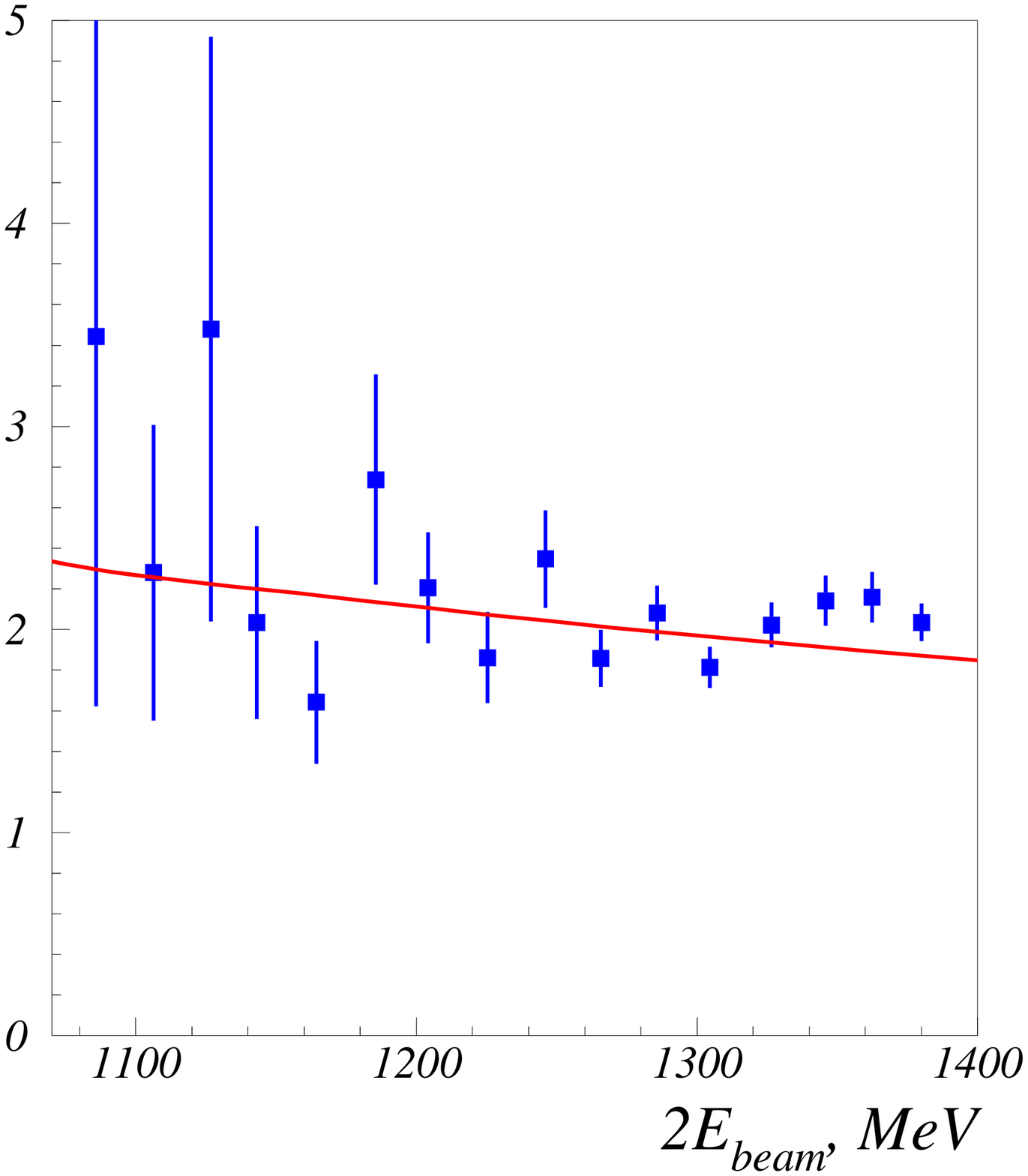}
\caption{\it
Energy dependence of 
$\sigma(\epm\to\pch)/\sigma(\epm\to\pne)$.
The contribution of $\omega\pi^{0}$ is subtracted.
The solid curve shows the theoretical prediction 
based on the $a_{1}\pi$ dominance
}
\label{xsratio}
\end{figure}

To obtain the $a_1\pi$ contribution to
the total $\pne$ cross section, the contribution
of the $\omega\pi$ intermediate state was subtracted. 
Such a procedure is possible because the interference between
$\omega\pi$ and $a_1\pi$ is numerically small ($\sim 5\%$) due to the
small width of  $\omega$ meson.
Figure~\ref{xsratio} presents the ratio of the cross sections 
$\sigma(\epm \to \pch)$ and $\sigma(\epm \to \pne)$ 
where the contribution of $\omega\pi^{0}$ is subtracted.
The solid curve shows the theoretical prediction 
based on the $a_{1}\pi$ dominance.

\begin{figure}[ht]
\vspace{9.0cm}
\includegraphics{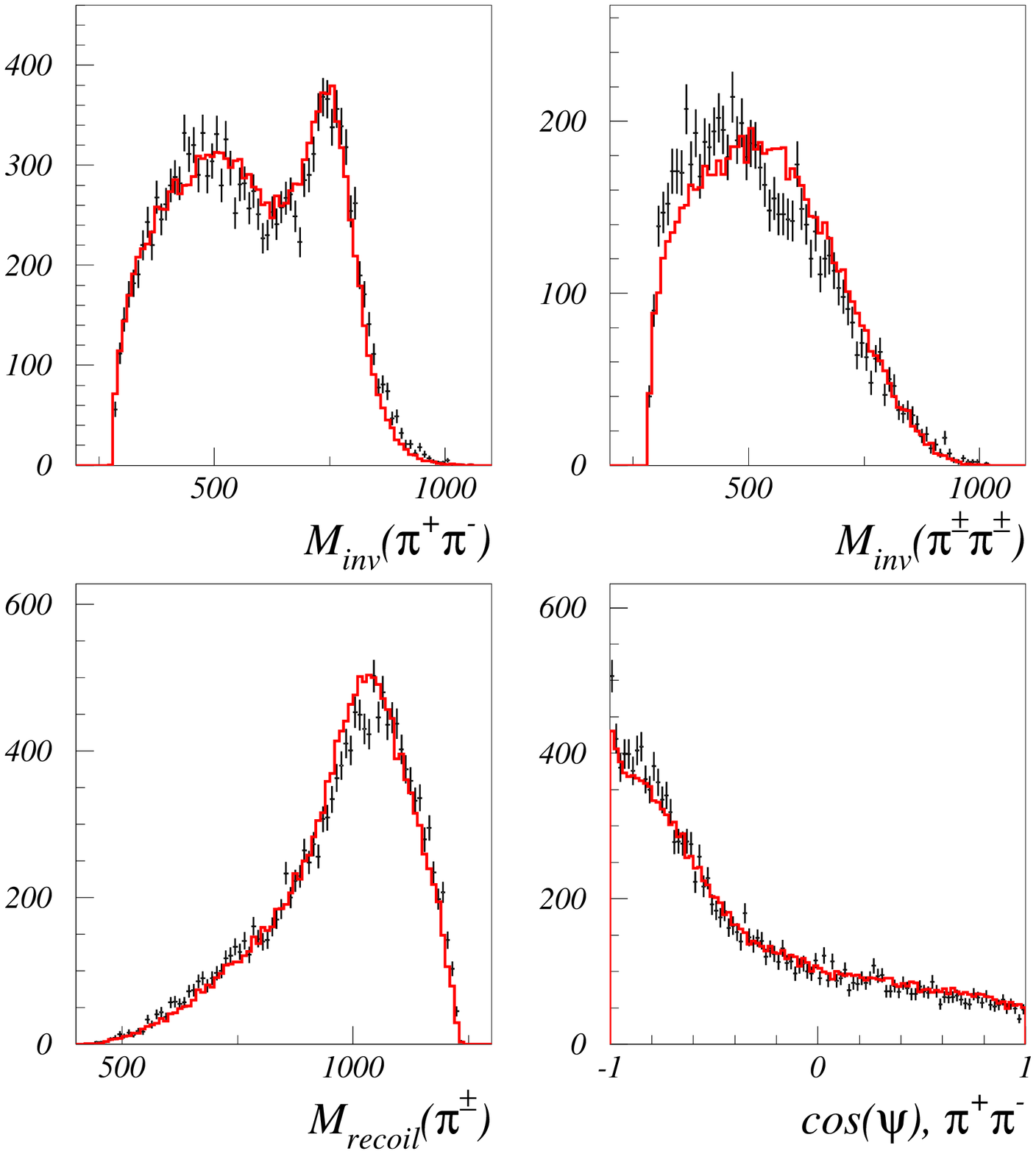}
\caption{\it
Distributions over
$M_{inv}(\pi^{+}\pi^{-})$,
$M_{inv}(\pi^{\pm}\pi^{\pm})$,
$M_{recoil}(\pi^{\pm})$ and
$cos(\psi_{\pi^{+}\pi^{-}})$
for $\pch$ events
}
\label{pi4slide}
\end{figure}

Figure~\ref{pi4slide} shows distributions over 
$M_{inv}(\pi^{+}\pi^{-})$,
$M_{inv}(\pi^{\pm}\pi^{\pm})$,
$M_{recoil}(\pi^{\pm})$ and
$cos(\psi_{\pi^{+}\pi^{-}})$ for \pch case.
One can see that the hypothesis of the $a_{1}\pi$ dominance is in
agreement with the data.

For the analysis of $\pne$ channel the data sample was subdivided into
two classes: 1. 
$min(|M_{recoil}(\pi^{0})-M_{\omega}|)~<~70~ MeV$,\\
2. $min(|M_{recoil}(\pi^{0})-M_{\omega}|)~>~70~ MeV$,
where $M_{\omega}$ is the $\omega$ mass.
The first class contains mostly $\omega\pi$ events while their
admixture in the second class is relatively small, about $(1\div5)\%$ 
depending on the beam energy.
In \cite{our} it was shown  that the process $\pne$ 
is well described in the minimal model in which there are
two intermediate states $\omega\pi$ and $a_1\pi$ only.
Similar consistence is observed at other energies.

\begin{figure}[ht]
\vspace{9.0cm}
\includegraphics{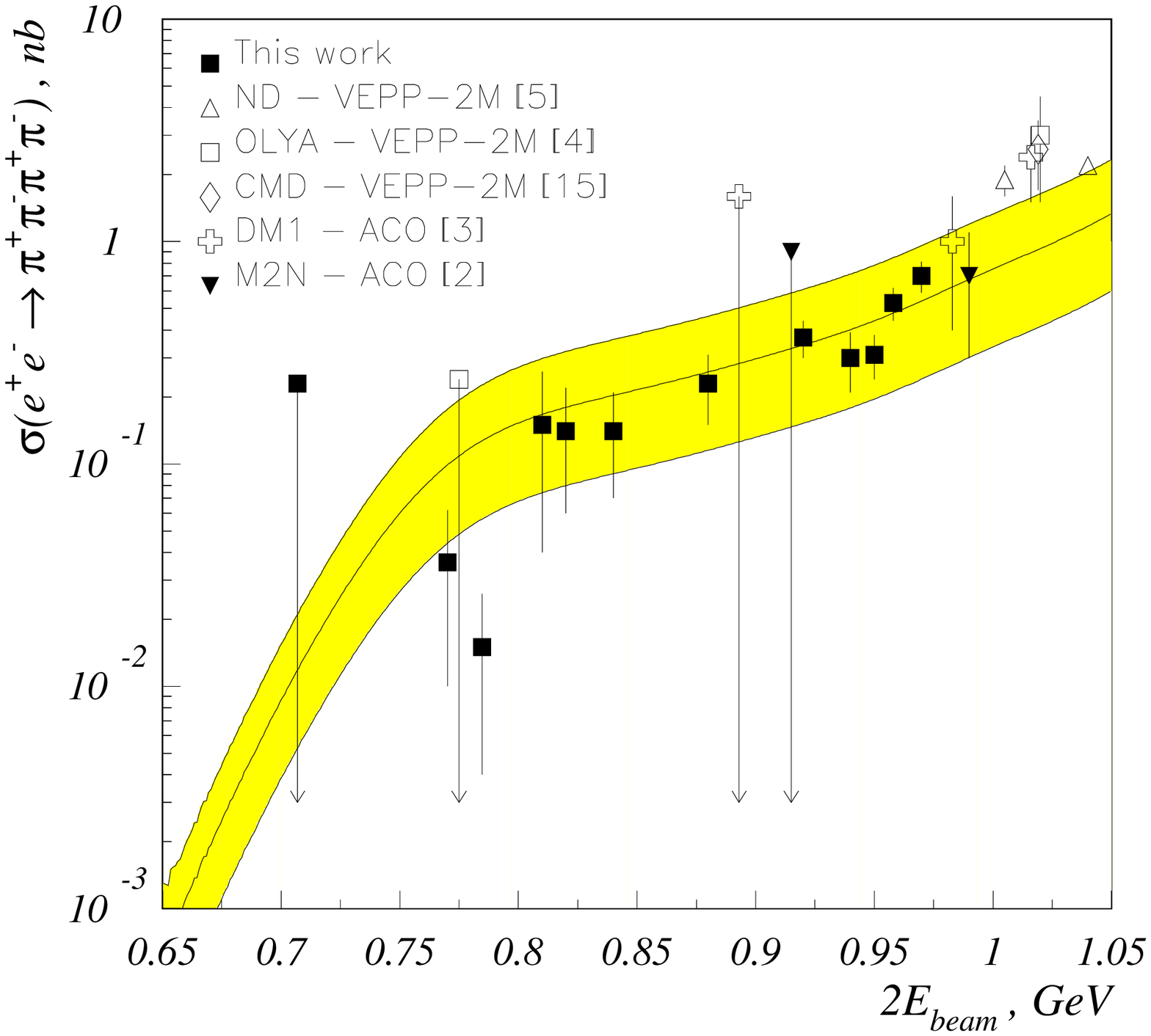}
\caption{\it Cross section of the process $e^+e^-\to\pch$
      below 1.05 GeV}
    \label{fig:xsec}
\end{figure}

Recently, using 3.07 $\mbox{pb}^{-1}$ of data collected
in the energy range 0.60--0.97 GeV by CMD-2, about 150 events
of the process $e^+\e^- \to \pch$ have been selected \cite{our2}. 
Figure~\ref{fig:xsec} shows the energy
dependence of the cross section below 1.05 GeV. 
For illustration, results from other measurements
\cite{cmd,nd,dm1,olya2,m2n} are  also demonstrated. 
The values of the cross section obtained in \cite{our2}  
are consistent with them and match the measurements of CMD-2
above the $\phi$ meson.
The overall systematic uncertainty was
estimated to be $\approx 12\%$.

The shaded area in Figure \ref{fig:xsec}
corresponds to the 
extrapolation of the energy dependence of the cross section from the
energy region above 1.05 GeV \cite{our}. 
The calculation assumed that the cross section behaviour is determined
by two interfering resonances - $\rho$ and its excitation $\rho'$ 
decaying into the final four pion state via the  
$a_1\pi$ intermediate mechanism. 
The central curve corresponds to the $a_1$ width of 600 MeV
optimal in the analysis whereas 
the upper and lower curves are obtained for the widths of
800 and 400 MeV respectively. 
It can be seen that the energy dependence of the data 
is consistent with the assumption of the $a_1\pi$ dominance
earlier established at higher energies \cite{our}. 

\section{$a_1\,\pi$ contribution to $\tau\to 4\pi\nu_{\tau}$ decay}

The initial hadron state  which
decays into  four pions for both $\tau$ decays and $e^+e^-$ annihilation
has the $\rho$-meson quantum numbers and is referred to as $\tilde\rho$.
Due to the conservation of vector current the probability $d\Gamma_1$
and $d\Gamma_2$ of $\tau^-\to\pi^-\pi^+\pi^-\pi^0\nu_\tau$ and
$\tau^-\to\pi^-\pi^0\pi^0\pi^0\nu_\tau$ decays respectively can be
written as
\beq\label{G1}
\frac{d\Gamma_i}{ds}=\frac{G^2|V_{ud}|^2}{96\pi^3 m_\tau^3}
(m_\tau^2+2s)(m_\tau^2-s)^2 R_{4\pi}\frac{dW_i}{W_1+W_2}
\eeq
where $G$ is the Fermi constant, $R_{4\pi}$ is the ratio of the cross
section $\epm\to 4\pi$ and 
$\epm\to \mu^+\mu^-$, $dW_1$ and $dW_2$ are the probabilities of
$\tilde\rho^-$ decays into $\pi^-\pi^+\pi^-\pi^0$ and
$\pi^-\pi^0\pi^0\pi^0$, respectively. Let $dW_3$ and $dW_4$ are
the probabilities of $\tilde\rho^0$ decays into $\pi^+\pi^-\pi^+\pi^-$ and
$\pi^+\pi^-\pi^0\pi^0$. Due to the isospin invariance, we have
$W_1=W_3/2+W_4$ and $W_2=W_3/2$ .
The explicit forms of the matrix elements, corresponding to $W_i$, are
presented in \cite{our1}. In order to get the predictions for $\tau$
decay  the interference between $\omega\pi$ amplitude and
$a_1\pi$ was neglected, and  \eq{G1} was written in the following form:
\beq\label{G11}
\frac{d\Gamma_1}{ds}=\frac{G^2|V_{ud}|^2}{96\pi^3 m_\tau^3}
(m_\tau^2+2s)(m_\tau^2-s)^2 \left[
 R_{\omega\pi}\frac{dW_{\omega}}{W_\omega} +
  R_{2\pi^+2\pi^-}\frac{dW_1}{W_3}\right] ,
\eeq
where $dW_{\omega}$ is the probability of $\tilde\rho^-\to\omega\pi^-$
decay, $R_{\omega\pi}$ is the  ratio of the cross section
$\epm\to \omega\pi$ and $\epm\to \mu^+\mu^-$.

In order to fix the parameters of the model the data of
 $e^+e^-\to 4\pi$ \cite{our} and $\tau^-\to 2\pi^0\pi^-2\nu_\tau$
\cite{3pi} were used. The mass of $a_1$ was taken from
the PDG table \cite{pdg} and the width was obtained as a result of
optimal description  of three pion
invariant mass distribution in $\tau^-\to 2\pi^0\pi^-\nu_\tau$ decay.
This value of $a_1$ width also provides a good description of
$e^+e^-\to 4\pi$ data. 

In \cite{3pi} it was obtained the evidence that  $a_1$ meson has significant
probability to decay into three pions through $\sigma\pi$
intermediate state. The data analysis of  $e^+e^-\to 4\pi$ 
also confirmed this statement. In \cite{our1} the admixture of
$\sigma\pi$ to the $a_1$ decay amplitude  was taken into account, 
and the parameters of this admixture were extracted from $e^+e^-\to 4\pi$ data.

\begin{figure}[ht]
\vspace{9.0cm}
\includegraphics{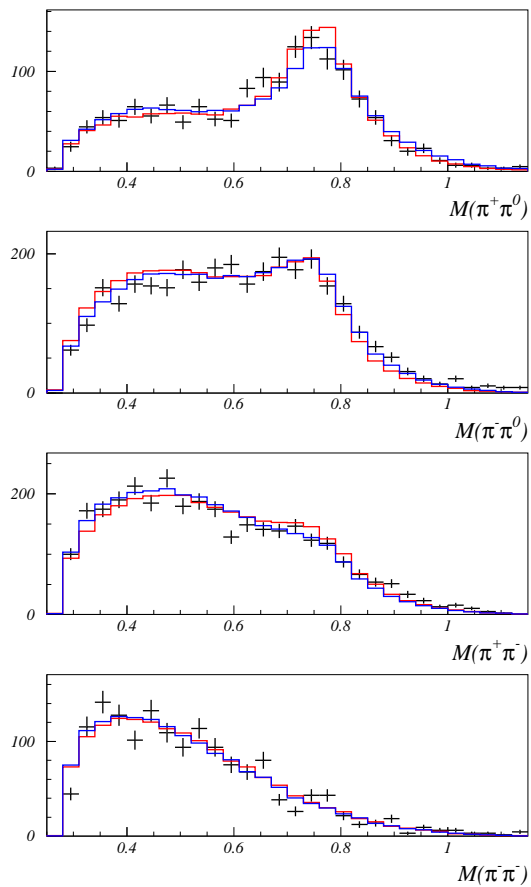}
\caption{\it
Two-pion invariant  mass distributions
for  $\tau^-\to 2\pi^-\pi^+\pi^0\nu_\tau$ decay after $\omega\pi^-$
events subtraction, obtained by ALEPH.
}
\label{aleph_pic}
\end{figure}

The most interesting information on the mechanism of four pion channel can be
obtained from two-pion mass distributions. 
In \cite{our1} it was shown  that data of CLEOII detector \cite{Cleo1} obtained
without subtraction of $\omega\pi^-$ contribution are in 
good agreement with predictions. 
The data obtained with the subtraction of $\omega\pi^-$ contribution allows
one to make a more detail comparison of the differential distributions
predicted within the assumption on $a_1\pi$ dominance. For this purpose
the data obtained by ALEPH \cite{Aleph} (see Fig.~\ref{aleph_pic}) and
very resent high-statistics data of CLEOII \cite{Cleo2} were used.
In the first case we see a good agreement in spite of  the absence of
possibility to take into account the detector efficiency and energy
resolution. The agreement for the data of CLEOII\cite{Cleo2}
is a little bit worse. Unfortunately,  in this data
the contributions of background events (such as $K^0\pi^-\pi^0\nu_\tau$
and $K^{\pm}\pi^{\mp}\pi^-\pi^0\nu_\tau$ ) were not subtracted,
though their fraction was significant (about 8\%). New analysis of
$\tau\to 4\pi\nu_{\tau}$ by CLEOII \cite{Cleonew} completely confirmed
conclusion on the production mechanism based on the assumption of
$a_1$ dominance.

\section*{Acknowledgements}
I am grateful to A.E.Bondar, S.I.Eidelman, N.I.Root, A.A.Salnikov, and
A.I.Sukhanov for useful discussions.

\end{document}